\documentclass[%
reprint,
superscriptaddress,
amsmath,amssymb,
aps,
prb,
floatfix,
a4paper
]{revtex4-2}

\usepackage{graphicx}
\usepackage{dcolumn}
\usepackage{bm}
\usepackage{hyperref}
\usepackage{gensymb}
\usepackage{xspace}
\usepackage{textcomp}
\usepackage{xcolor}

\def\cuf{Cu$_2$F$_5$\xspace}
\def\cufx{Cu$_2$F$_{5-x}$\xspace}

\begin{document}


\title{2D to 1D magnetic interactions evolution in Cu$_2$F$_{5-x}$ through electron doping by fluoride non-stoichiometry}

\author{Dmitry~M.~Korotin}
	\email{dmitry@korotin.name}
	\affiliation{M.N. Mikheev Institute of Metal Physics of Ural Branch of Russian Academy of Sciences, 18 S. Kovalevskaya St., Yekaterinburg, 620137, Russia.}

\author{Dmitry~Y.~Novoselov}
	\affiliation{M.N. Mikheev Institute of Metal Physics of Ural Branch of Russian Academy of Sciences, 18 S. Kovalevskaya St., Yekaterinburg, 620137, Russia.}
    \affiliation{Department of Theoretical Physics and Applied Mathematics, Ural Federal University, 19 Mira St., Yekaterinburg 620002, Russia}

\author{Vladimir~I.~Anisimov}
	\affiliation{M.N. Mikheev Institute of Metal Physics of Ural Branch of Russian Academy of Sciences, 18 S. Kovalevskaya St., Yekaterinburg, 620137, Russia.}
    \affiliation{Department of Theoretical Physics and Applied Mathematics, Ural Federal University, 19 Mira St., Yekaterinburg 620002, Russia}

\date{\today}

\begin{abstract}
The copper fluoride \cuf is a compound with 2D-magnetic exchange interactions between the Cu ions in the $S=1$ and $S=\frac{1}{2}$ spin-states. Using {\em ab-initio} calculations, we predict that the existence of 5\% vacancies in the fluoride sublattice of \cuf results in the drastic transformation of the spin-state of all copper ions and the final spin-states are $S=\frac{1}{2}$ and $S=0$. Consequently, the anisotropy of magnetic interactions increases, and the 1D linear chains of the Cu $d^9$, $S=\frac{1}{2}$ ions appear. We also propose a microscopic mechanism of such exchange interaction transformation via  CuO$_6$ octahedra elongation.
\end{abstract}

\maketitle

\section{Introduction}

Cuprates are well-known objects for the low-dimensional magnetism appearance.
With such structural building blocks as CuO$_6$ octahedra and CuO$_4$ plaquettes, and due to the presence of copper ion in $d^9$ electronic configuration, there is a variety of magnetic structures. 
In perovskite-like KCuF$_3$~\cite{PhysRevB.52.R5467,exchanges}, which structure is formed with corner-sharing CuF$_6$ octahedra, there is the G-type antiferromagnetic ordering of moments with existence of 1D magnetic chains of Cu ions.
 2D antiferromagnetic ladders are realized in Sr$_{n-1}$Cu$_{n+1}$O$_{2n}$~\cite{Gopalan1994} and SrCu$_2$O$_3$~\cite{Muller1998} with edge-sharing CuO$_4$ plaquettes. And the corner-sharing plaquettes in Sr$_2$CuO$_4$~\cite{Rosner1997} and AgCuVO$_4$~\cite{Moller2009} led to appearance of the one-dimensional chain of $S=\frac{1}{2}$ Cu ions.


The plethora of spin lattices mentioned above exist in the copper-oxygen complexes. 
The stability of the copper-fluoride complex, \cuf, was predicted recently~\cite{novelFluorides}. Structurally it is formed by both blocks: the octahedra (CuF$_6$) and the plaquettes (CuF$_4$). Consequently, one can expect that the magnetic anisotropy, similar to the one seen in the cuprates, could appear in the fluoride with or without additional doping with carriers. In our previous work~\cite{Korotin2021}, within the DFT+U calculations, we show that in stoichiometric \cuf, the Cu ions are in the $S=1$ and $S=\frac{1}{2}$ spin-states ($d^8$ and $d^9$ electronic configuration). Additionally, we obtained that the Heisenberg exchange interaction along the $a$ crystal axis is tiny. 
In the (100)-plane there is antiferromagnetic superexchange between the half-filled $d$-orbitals of the nearest Cu ions through fluoride $p$-states located in between.
As a consequence, the 2D spin-lattice exists in stoichiometric \cuf. 

One can assume that if an extra electron occupies the half-filled $d$-orbital of Cu ion in \cuf, the corresponding Cu-$d$ $\leftrightarrow$ F-$p$ $\leftrightarrow$ Cu-$d$ superexchange interaction will be destroyed. More extra electrons will cause a significant modification of the exchange interactions.
Our purpose was to find conditions under which the one-dimensional magnetic interactions prevail in copper fluoride.

We have produced additional electrons in the cell considering the doped compound -- \cufx. Each fluoride vacancy results in an extra electron that occupies one of the copper $d$-orbitals. As is shown in the following sections, not only electronic and magnetic, but the crystal structure of the compound also evolves following the doping.

\section{Methods}\label{sec:methods}
In this paper, we follow the methodology defined in the previous one, describing \cuf~\cite{Korotin2021}.
All calculations were performed using Quantum-ESPRESSO~\cite{Giannozzi2009} package with pseudopotentials from pslibrary set~\cite{pslibrary}. The exchange-correlation functional was chosen to be in Perdew-Burke-Ernzerhof~\cite{PBE} form. The energy cutoff for plane wave wave functions and charge density expansion was set to 50~Ry and 400~Ry, respectively. Integration in the reciprocal space was done on a regular $8\times8\times8$ $k$-points mesh in the irreducible part of the Brillouin zone. 

Electronic correlations were treated within the DFT+U method~\cite{Cococcioni2005} with the Hubbard $U$ value equals 6~eV. 
As it was shown for the parent compound \cuf, even the $U$ value of 4 eV is enough for the band gap to appear. The variation of the $U$ from 4 to 8~eV doesn't change the electronic structure qualitatively and affects the Heisenberg exchange interaction parameters only slightly.
The Hund parameter $J$ = 0.9~eV was set to its typical value for cuprates~\cite{Blaha2005,Leonov2008}.

The convergence criteria used for crystal cell relaxation within DFT+U are: total energy $< 10^{-6}$~Ry, total force $< 10^{-3}$~Ry/Bohr, pressure $< 0.1$~kbar.

\section{\label{sec:results} Results }

\begin{figure}
	\includegraphics[width=\columnwidth]{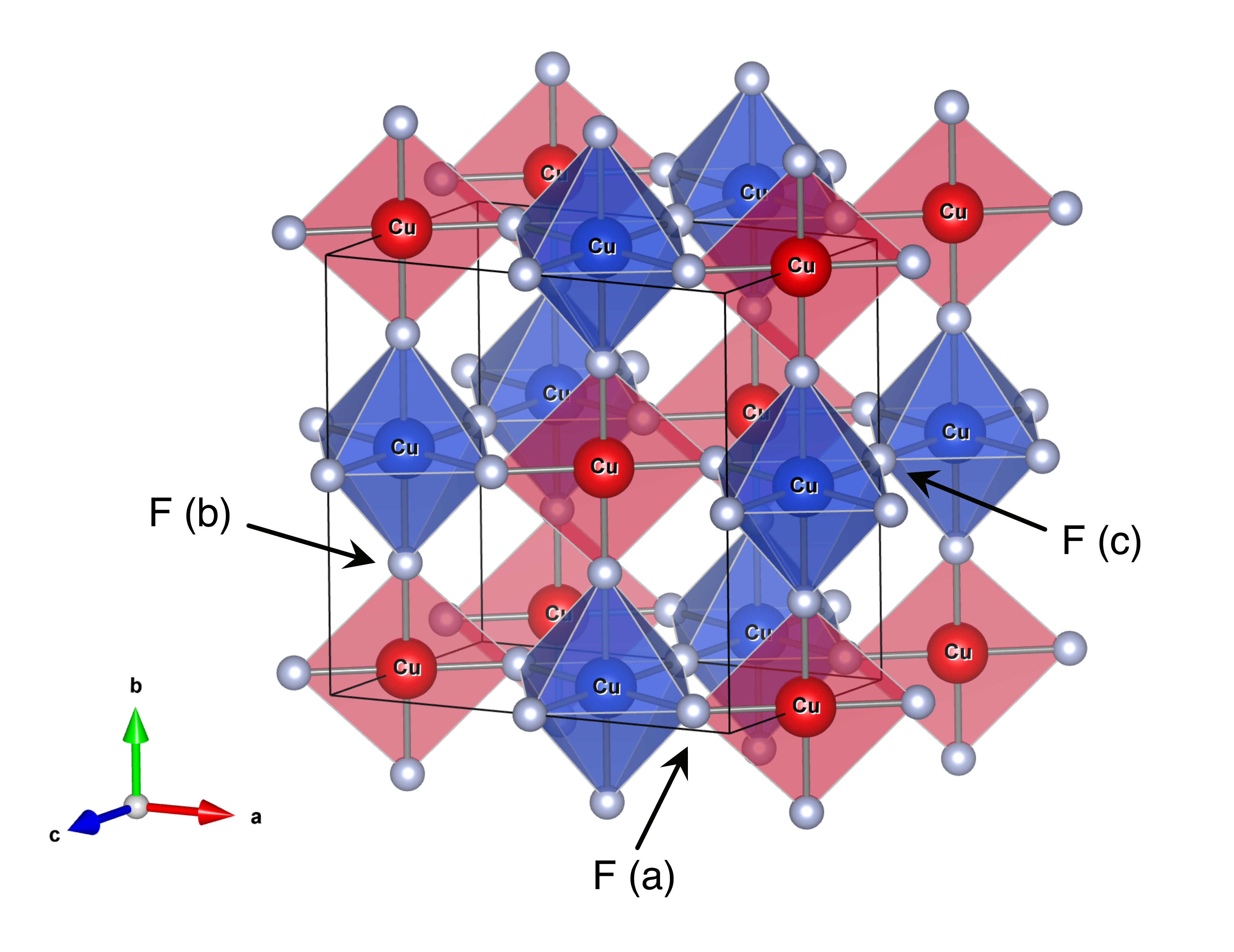}
	\caption{Crystal structure of \cuf and the three types of fluoride ions that were removed to obtain the structure of \cufx. Blue spheres denote Cu ions inside the ligand's octahedron, red spheres denote Cu ions in the center of plaquettes, and gray spheres -- F ions. Visualized using VESTA~\cite{VESTA}.}
	\label{fig:structure}
\end{figure}

There are three different structural sites for the fluoride ion in Cu$_2$F$_5$ that could be substituted by a vacancy: (a) the fluoride shared by CuF$_4$ plaquette and CuF$_6$ octahedron that are placed along the \emph{a} axis of the cell; (b) the ion shared by CuF$_4$ plaquette and CuF$_6$ octahedron in the direction of the \emph{b} axis; (c) the fluoride belonging to the two CuF$_6$ octahedra along the \emph{c} axis. The Cu-F-Cu bond angle is 180 for the (b) and (c) ions and only 129 degrees for the case-(a) fluoride ion, which results in much weaker magnetic exchange interaction in the direction of the \emph{a} axis. The absence of one of these F-ions would naturally lead to a change in the electronic configuration of the nearest Cu ions. With the destruction of the Cu-F-Cu superexchange path, the disappearance of one fluoride ion probably will change the magnetic exchange interaction pattern in Cu$_2$F$_5$.

In the Cu$_2$F$_5$ crystal cell containing 8 Cu ions (see Fig.~\ref{fig:structure}), we removed the F ion in (a), (b), or (c) position sequentially, as described above, and performed full cell relaxation within the DFT+U approach to obtain a ground state crystal structure. The used cell size corresponds to a 5\% concentration of fluoride vacancies (Cu$_2$F$_{4.75}$), and we are focused on this simplest and the most visual case.

After the relaxation, we compared the enthalpies of the obtained structures. The lowest enthalpy has the cell where the vacancy is placed instead of the F ion in the (b) position. If F is removed from the (c) position, the enthalpy is +45 meV / formula unit higher and it is +46 meV/formula unit higher for the vacancy in the (a) position. We conclude that the favorable vacancy localization site in Cu$_2$F$_5$ is the F ion shared by CuF$_4$ plaquette and CuF$_6$ octahedron in the direction of the \emph{b} axis. Below under the Cu$_2$F$_{4.75}$ or Cu$_2$F$_{5-x}$, we mean the corresponding crystal structure.

\begin{table}[t!]
	\centering
	 \begin{tabular}{c | c | c} 
	 \hline\hline
	 Parameter & \cuf & Cu$_2$F$_{4.75}$ \\ 
	 \hline\hline
	 Cell volume (\AA$^3$) & 369.3 & 393.6  \\ 
	 Cu$_{octa}$ - Cu$_{octa}$ distance along $a$ (\AA) & 6.98 & 6.96 \\
	 Cu$_{octa}$ - Cu$_{octa}$ distance along $b$ (\AA) & 7.61 & 8.08 \\
	 Cu$_{octa}$ - Cu$_{octa}$ distance along $c$ (\AA) & 7.60 & 7.52 \\
	 Average Cu$_{plaq}$-F bond length (\AA) & 1.90 & 1.80 \\
	 Average Cu$_{octa}$-F bond length (\AA) & 1.93 & 2.04 \\
	 Average Cu$_{octa}$-F distance along $b$ (\AA) & 1.93 & 2.26 \\
	 \hline\hline
	\end{tabular}
	\caption{Transformation of crystal structure of \cuf with electrons doping via vacancies.}
	\label{tab:distortions}
	\end{table}

The vacancy creation results in the crystal structure distortions presented in Table~\ref{tab:distortions}. The unit cell volume increase of 6.6\% happens from the elongation of the cell along the $b$ lattice vector. At the same time, there is an expansion of the CuF$_6$ octahedra and a decrease of the average Cu-F bond length within the CuF$_4$ plaquette in Cu$_2$F$_{5-x}$.

The two Cu ions, that had the F ion in between in Cu$_2$F$_5$, and have the vacancy site instead in Cu$_2$F$_{4.75}$, are slightly shifted to each other along the \emph{b} crystal axis. Hereafter, the Cu-F bonds of the \emph{ex}-CuF$_4$ plaquette and the \emph{ex}-CuF$_6$ octahedron are distorted when they lose the shared fluorine ion. For the next-nearest to the vacancy Cu-F octahedra and plaquettes, one can still say that the local environment for the Cu ion remains an octahedron and a plaquette. The structure files for Cu$_2$F$_5$ and Cu$_2$F$_{4.75}$ could be found here~\cite{cryst_struc_dataset}.

\begin{figure}
	\includegraphics[width=\columnwidth]{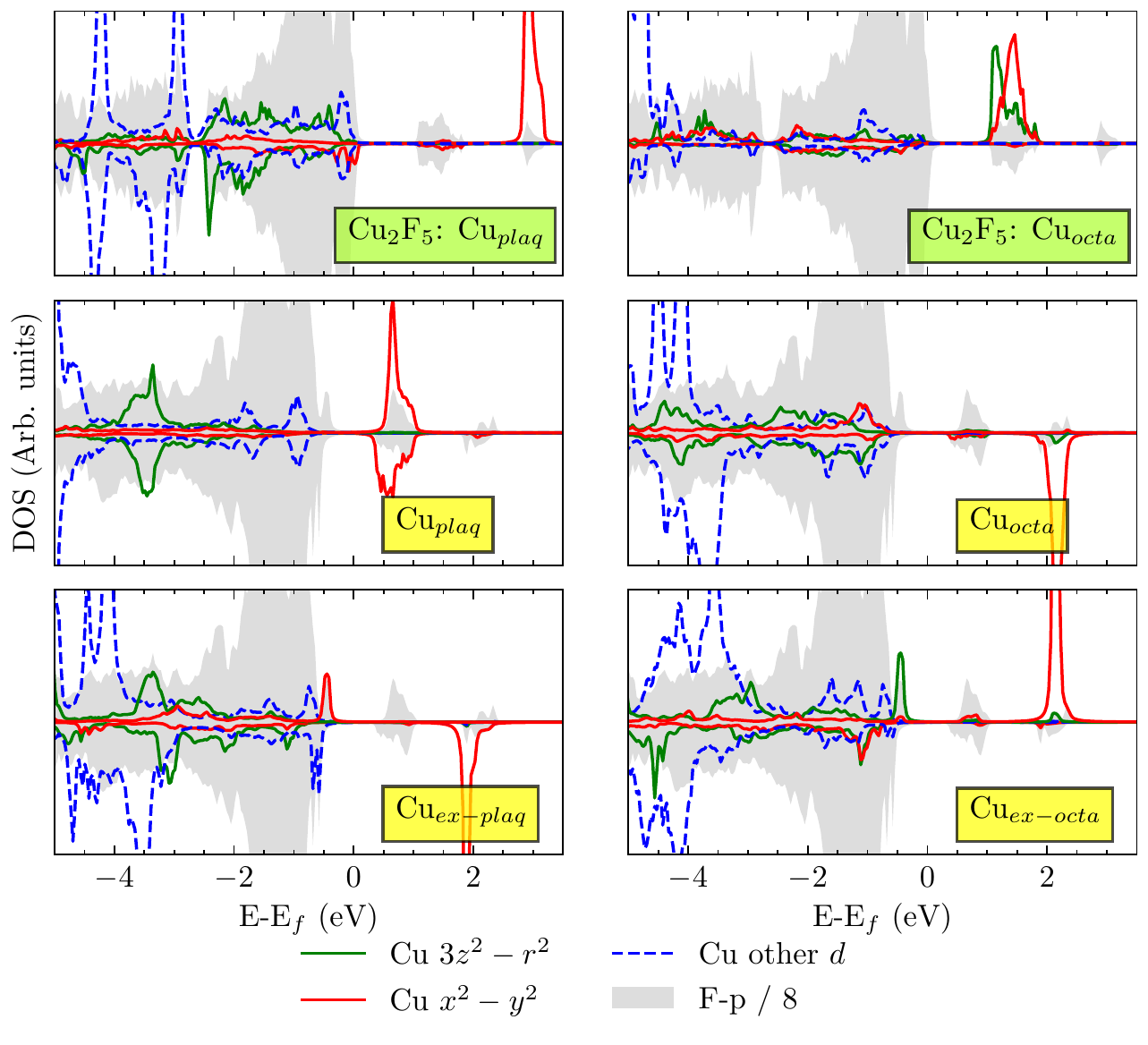}
	\caption{Partial densities of states for \cuf (upper panel) and \cufx (middle and lower panels). Positive/negative pDOSes correspond to spin-up/down states respectively.}
	\label{fig:pdoses}
\end{figure}

The Cu$_2$F$_{4.75}$ is an antiferromagnetic insulator with a 0.91 eV band gap. The calculated partial densities of states (pDOS) are presented in Fig.~\ref{fig:pdoses}. A plausible assumption would be to say that all copper ions inside the fluoride octahedra have very similar pDOSes despite the various distortions of CuF$_6$ structures regarding the distance from the vacancy. The same is true for the CuF$_4$ plaquettes. The corresponding pDOSes are marked in Fig.~\ref{fig:pdoses} as Cu$_{octa}$ and Cu$_{plaq}$ respectively. Here and below, we refer to the Cu $3z^2-r^2$, $x^2-y^2$, etc. orbitals in terms of the local coordinate system for each Cu ion, with the $z$ direction is perpendicular to the plaquette plane for the Cu$_{plaq}$ ion and the $z$ direction is along with the crystal $b$ vector for the Cu$_{octa}$ ion.

From the analysis of pDOS, one can see that Cu ions in the octahedral environment obtained an additional electron and became Cu$^{2+}$ ions in $d^9$, $S=\frac{1}{2}$ electronic configuration: all $t_{2g}$ and $3z^2-r^2$ orbitals are filled, and the $x^2-y^2$ orbital is half-filled. At the same time, the plaquette-surrounded Cu ions are now in $d^8$, $S=0$ configuration and have negligible magnetic moments. Both Cu$_{ex-plaq}$ and Cu$_{ex-octa}$ ions are in $d^9$, $S=\frac{1}{2}$ electronic configuration with the hole at the $x^2-y^2$ orbital. 
We started our calculation from the cell containing 4 Cu$_{octa}$ ions in $d^8$, $S=1$ configuration + 4 Cu$_{plaq}$ ions in $d^9$, $S=\frac{1}{2}$ configuration + an electron from the vacancy. At the end we've obtained 3 Cu$_{plaq}$ ions in $d^8$, $S=0$ + 5 Cu ions (3 Cu$_{octa}$, Cu$_{ex-plaq}$, Cu$_{ex-octa}$) in $d^9$, $S=\frac{1}{2}$ configuration.

In the Cu$_2$F$_5$~\cite{Korotin2021} the lower unoccupied state is the $3z^2-r^2$ orbital of the Cu$_{octa}$ ions (Fig.~\ref{fig:pdoses}, upper panel, right graph). We assumed that as a consequence of the doping, the additional electron will occupy some of these orbitals. Consequently, metallization of the Cu$_2$F$_5$ with electron doping was expected due to the appearance of the partially filled Cu$_{octa}$ $3z^2-r^2$ states.
Surprisingly we observe an electron transfer from the Cu$_{plaq}$ ions to the Cu$_{octa}$ ions in Cu$_2$F$_5$ with doping. We interpret it in the following way. Due to the partial occupation of the Cu$_{octa}$ $3z^2-r^2$ orbital, the corresponding octahedron elongates in the $c$ direction, that results in the expansion of the cell along the $c$ axis. Consequently, all the CuF$_6$ octahedra are elongated. Since the Cu-F distances in the $b$ direction within the octahedron become large, the Cu$_{octa}$ $3z^2-r^2$ orbital turns energetically more favorable than even the Cu$_{plaq}$ $x^2-y^2$ orbital. Electrons that occupied the Cu$_{plaq}$ $x^2-y^2$ state in the stoichiometric Cu$_2$F$_5$ leave it in Cu$_2$F$_{5-x}$ and fill the Cu$_{octa}$ ions d-shell. As a result, Cu$_{plaq}$ ions that have the $d^9$ configuration in Cu$_2$F$_5$, become $d^8$ in the doped structure, and Cu$_{octa}$ ions change their configuration in reverse way.

To confirm such evolution of crystal and electronic structure arises from the electronic degrees of freedom, not from interactions between the vacancy states, we modeled the electrons doping in a stoichiometric \cuf. The fluoride ion wasn't removed. We just added one extra electron in the cell and relaxed the crystal structure.

As a result, the same effect was qualitatively reproduced. The cell volume increase by 17.5\% with its significant elongation along $b$ direction (Cu$_{octa}$-Cu$_{octa}$ distance grows up to 8.08~\AA). The CuF$_6$ octahedra are stretched along $b$ direction too, with the corresponding average Cu$_{octa}$-F bond length equals 2.25~\AA. The spin-state of the copper ions was also changed as a result of the existence of additional electrons: the Cu$_{octa}$ ions became $d^9$, $S=\frac{1}{2}$.

\begin{figure}
	\includegraphics[width=0.6\columnwidth]{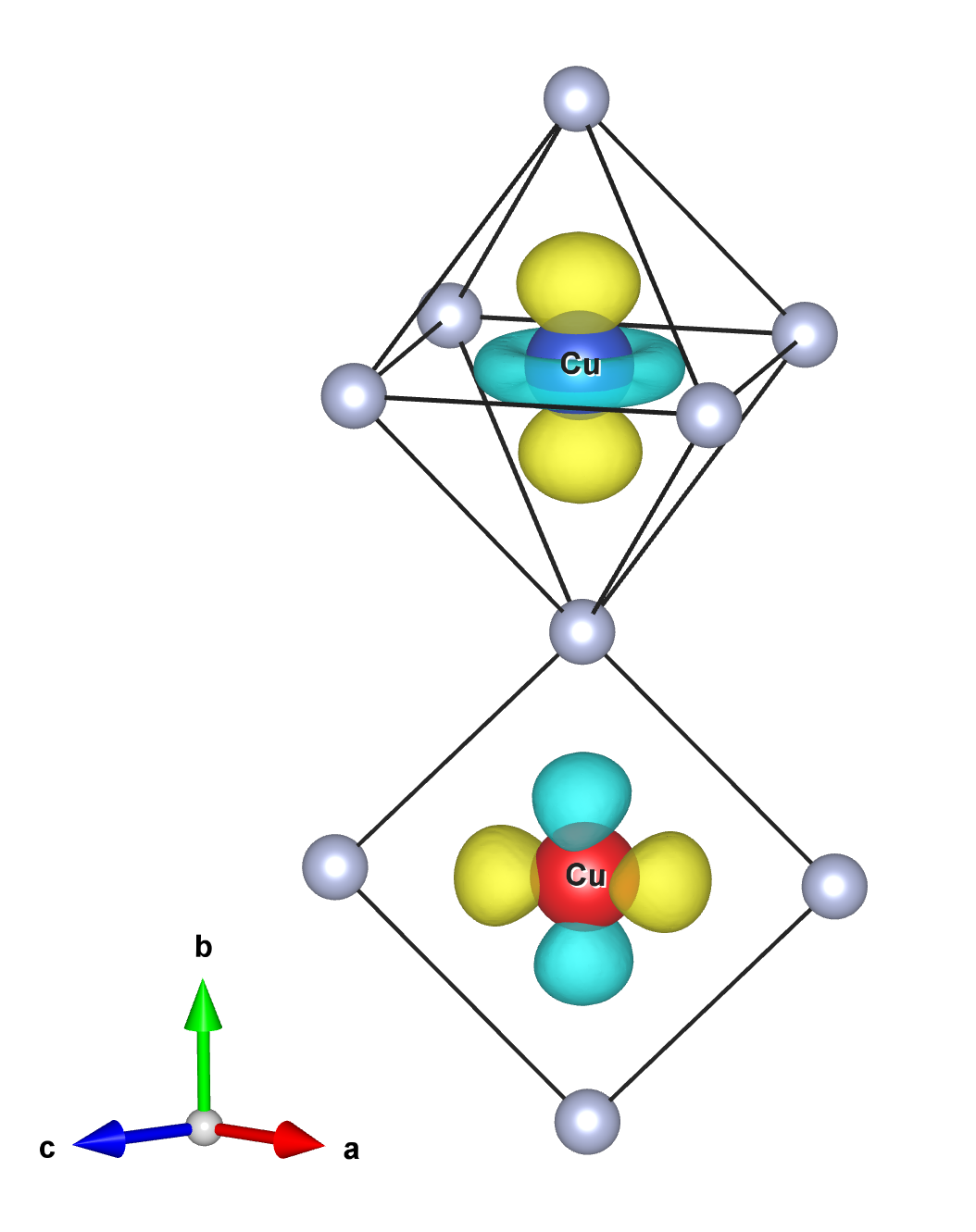}
	\caption{Two half-filled $d$-orbitals of Cu ions that provide the superexchange interaction via F $p$-orbital along the $b$ axis in stoichiometric \cuf. In \cufx the $d_{3z^2-r^2}$ become fully filled and the superexchange is suppressed.}
	\label{fig:orbitals}
\end{figure}

We concluded that the electronic configuration and spin state of the copper ions in \cuf evolve in the same manner if an extra electron appears regardless of the origin of the extra electron (vacancy or manual increase in the number of carriers within the cell). 
We continue our presentation and reasoning below for \cufx with the vacancy.

The half-filled $d_{3z^2-r^2}$-orbital of Cu$_{octa}$ ions in \cuf gives the superexchange interaction path between Cu$_{octa}$ and Cu$_{plaq}$ ions along the $b$ crystal axis as it is shown in Figure~\ref{fig:orbitals}. The electrons hoping between  Cu$_{octa}$ $d_{3z^2-r^2}$ and Cu$_{plaq}$ $d_{x^2-^y2}$ via the F-$p$ orbital in between led to the antiferromagnetic exchange.
In \cufx the possibility of such a hoping of electrons is suppressed, since Cu$_{octa}$ $d_{3z^2-r^2}$ becomes filled. Consequently, the corresponding superexchange interaction along the $b$ crystal axis vanishes.

The described evolution of electronic structure has an outcome that the 1D chains of Cu$_{octa}$ ions with a hole on the $d_{x^2-y^2}$-orbital appear. Taking into account the existence of the fluoride ion between such copper ions and following the Goodenough-Kanamori rule~\cite{Goodenough2008}, the antiferromagnetic superexchange interaction will emerge along the $c$ crystal axis.

\begin{figure*}[t!]
	\includegraphics[width=0.9\textwidth]{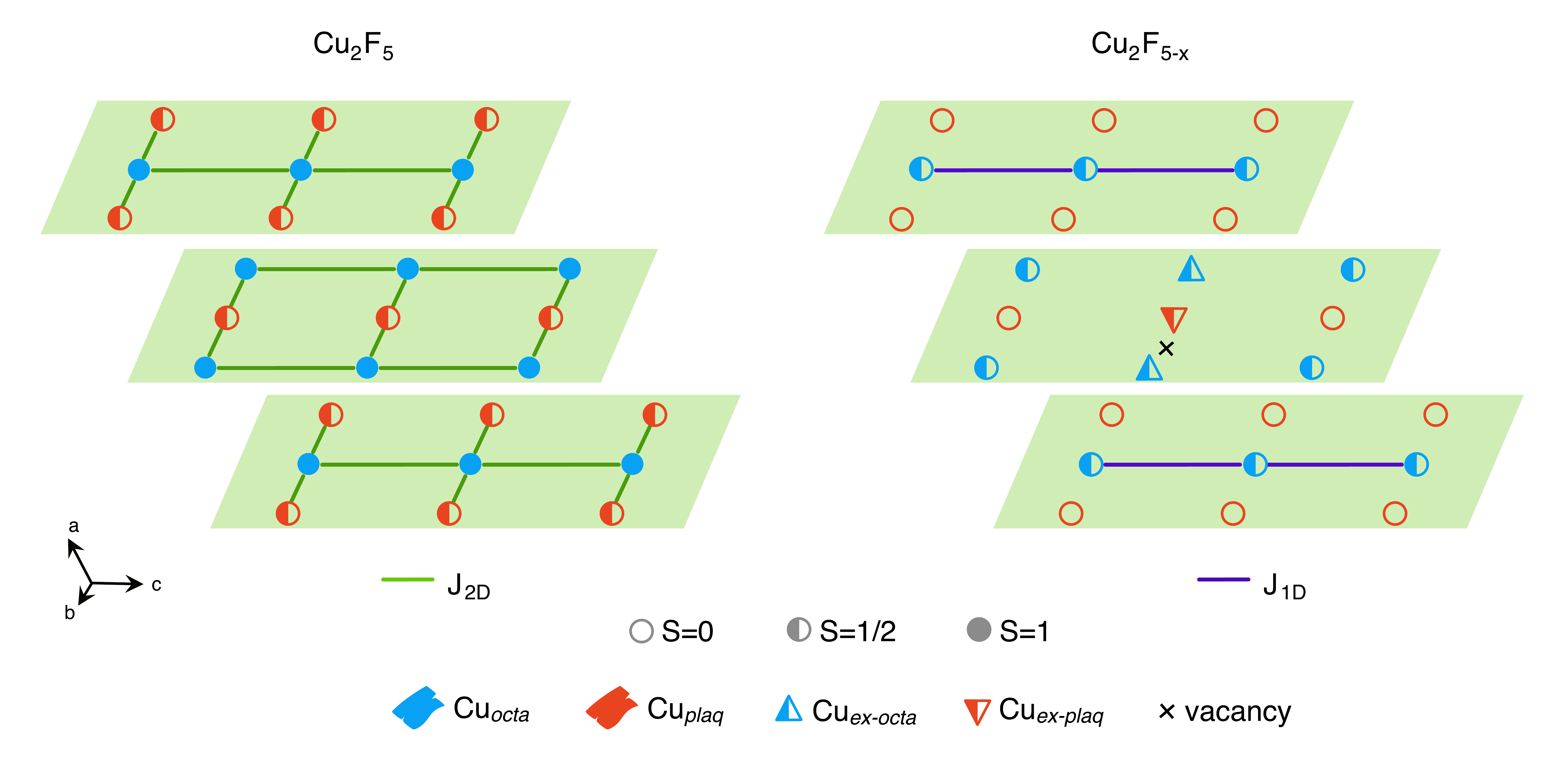}
	\caption{Evolution of spin-states and patterns of the exchange interaction between the Cu ion from \cuf (left) to \cufx (right). Blue elements denote Cu ions in the octahedral surrounding, red elements denote Cu ions in the center of fluoride plaquettes. 
	The blue and red half-filled triangles correspond to Cu$_{ex-octa}$, $S=\frac{1}{2}$ and Cu$_{ex-plaq}$, $S=\frac{1}{2}$ ions respectively. The filled, half-filled and empty circles denote $S=1$, $S=\frac{1}{2}$, and $S=0$ spin-states of the ions.
	The (100) lattice planes are shown with a light green color for an eye guide. The vacancy position is shown with a black cross.
	The strongest exchange interactions are $J_{2D}$ in \cuf (green line) and $J_{1D}$ in \cufx (violet line). Other exchange interactions are negligible. Fluorine ions are not shown for clarity.}
	\label{fig:exchanges}
\end{figure*}

Using Green's function method based on magnetic-force linear response theory~\cite{exchanges}, we computed the Heisenberg exchange interaction between Cu ions up to the 9th nearest neighbor. 
The model Hamiltonian has the form:
$H = -\sum_{\langle ij\rangle} J_{ij} {\bf e}_i {\bf e}_j,$
where ${\bf e}_i$ are the unit vectors pointing in the direction of the $i$th site magnetization, and the summation runs once over each ion pair. 

Only one exchange interaction survived under doping. The antiferromagnetic exchange between the Cu$_{octa}$ ions along the $c$-axis is $J_{1D}$ = -29.5~meV. The second largest interaction $\approx$ -3.3~meV is between Cu$_{octa}$ and Cu$_{ex-plaq}$ ions along the [101]-direction.
The pattern of the 1D magnetic chains stems from such exchanges with the interchain interaction being an order of magnitude smaller than the intrachain one. It is shown in Fig.~\ref{fig:exchanges}, right panel.
The figure illustrates also the evolution of the strongest exchange interactions in \cuf that arose from electron doping. The absolute value of the exchange remains almost the same:$J_{2D} \approx -33~\textrm{meV}$ in \cuf, but the dimension of the interaction decreases from 2D to 1D as a result of doping.

We point here to an analogy to the exchange interaction pattern that exists in KCuF$3$. The potassium copper fluoride is formed with the CuF$_6$ octahedra, and all copper ions have a hole on the $d_{x^2-y^2}$ orbital~\cite{Leonov2008}. Since there is the Jahn-Teller distortions of the octahedra, the lobes of the half-empty $d$-orbitals of the nearest Cu ions are perpendicular to each other in the [001]-plane. Consequently, the superexchange via the F $p$-orbitals in the [001]-plane is negligible in KCuF$_3$ and the 1D chains of antiferromagnetically ordered moments appear along the $c$ crystal axis. Therefore, depsite the different building blocks of the structure in \cufx (octahedra and plaquettes) and KCuF$_3$ (octahedra only), there is a similarity in the magnetic interactions picture between these two compounds.

\section{Conclusion}

Using the DFT+U calculations, we explored the influence of the fluoride vacancy appearance on the crystal, electronic and magnetic structure of \cufx. Extra electrons, which resulted from the absence of the fluorine ion, are shown to result in the elongation of the CuF$_6$ octahedra along the $b$ crystal axis and then the $3z^2-r^2$ orbital of all Cu$_{octa}$ ions becomes occupied. As a result, all the Cu ions in the center of CuF$_6$ octahedra get $S=\frac{1}{2}$ spin configuration, and the Cu ions inside the CuF$_4$ plaquettes become non-magnetic ($S=0$). Such significant effect become apparent even when one extra electron per 4x formula unit is added. 
The antiferromagnetic linear chains of copper ions appear along the $c$-axis of the crystal. The interchain exchange interaction is ten times smaller than the largest intrachain one.
Our calculations show consistently that \cufx can be described as a quasi-one-dimensional $S=\frac{1}{2}$ Heisenberg chain in a good approximation.

\section*{Acknowledgments}
Calculation of the ground state crystal structure for the doped \cufx was carried out within the state assignment of Ministry of Science and Higher Education of the Russian Federation (theme “Electron” No. 122021000039-4). Results on the spin-lattice evolution with doping were obtained with the support of the Russian Science Foundation (project No. 19-12-00012).

\bibliographystyle{apsrev4-1}
\bibliography{main}

\end{document}